\documentclass{svjour2}                    
\smartqed  
\usepackage{graphicx}
\journalname{}
\begin{document}

\title{Tops and Writhing DNA
}


\author{Joseph Samuel         \and
        Supurna Sinha 
}


\institute{J. Samuel \at
              Raman Research Institute \\
              Tel.: +91-80-2361 0122\\
              Fax: +91-80-2361 0492\\
              \email{sam@rri.res.in}           
           \and
           S. Sinha \at
              Raman Research Institute
}

\date{}

\maketitle

\begin{abstract}
The torsional elasticity of semiflexible polymers like DNA
is of biological significance. A mathematical treatment of this
problem was begun by Fuller using the relation between link, twist and
writhe, but progress has been hindered by the non-local nature of the
writhe. This stands in the way of
an analytic
statistical mechanical treatment, which takes into account thermal
fluctuations, in computing the partition function.
In this paper we use the well known analogy with the dynamics of tops to
show that when subjected to stretch and twist,
the polymer configurations which dominate the partition function
admit a local writhe formulation in the
spirit of Fuller and thus provide an underlying justification for the use
of Fuller's
``local writhe expression''
which leads to considerable mathematical simplification
in solving theoretical models of DNA and elucidating
their predictions.
Our result facilitates comparison of the theoretical models with single 
molecule
micromanipulation experiments and computer simulations.

\keywords{tops \and thermal fluctuations \and writhe \and torsional elasticity of polymers}
\PACS{64.70.qd\and 82.37.Rs\and45.20.da\and 
82.39.Pj\and45.20.Jj\and 87.14.g}
\end{abstract}

\section{Introduction}
\label{intro}
DNA, which carries the genetic code of living organisms is
a semiflexible polymer. The elastic properties of DNA are relevant
to a number of biological processes\cite{wang}.
Twist rigidity plays an
important role in packaging metres of DNA efficiently
in the tiny volume of the cell nucleus, just a
few microns
across. This involves DNA-histone
association which makes use of supercoiling in an
essential way.
The process of DNA
transcription can generate and be regulated by supercoiling\cite{strick}.

To
understand
such
effects, there have been single molecule experiments\cite{strick} which
pull and
twist
DNA molecules to probe their elastic properties. However, the 
interpretation
of these experiments has been hindered by a lack of
understanding of the geometry of writhe,   
the twisting of a DNA backbone.
The writhe of a DNA
polymer configuration, viewed as a space curve, is
a nonlocal quantity which has subtle geometric and topological
properties. The
nonlocality of the writhe
makes it cumbersome  to use in analytical or computational work
\cite{markotwist,volo,maggs,roseu,swig,berpri,canturck}.
This paper is devoted to
understanding the writhe for configurations
which dominate the partition function: those close to the minima
of the energy functional. More precisely, we show that for the stable
solutions of the Euler-Lagrange equations (local minima of the energy)
the writhe can be expressed as a local integral. This result can be used
to develop an analytical approach to the computation of the partition
function.

The most popular theoretical model for describing semiflexible
polymers is the worm-like-chain (WLC)\cite{marko}. The WLC is the ``
harmonic
oscillator'' in the field of semiflexible polymers: it is a simple model
which deftly captures much of the physics.
Given the importance of semiflexible polymers to biology,
from lipids to cytoskeletons to DNA,
the WLC deserves to be much better understood than it presently is.  
Two extreme situations are relatively well understood. If the polymer is
highly stretched and
nearly straight, one can use perturbation theory about the straight line
to calculate its elastic
properties\cite{nelson,sinha,freesinha,samsupabhi,abhi}.
Treatments also exist \cite{volo}
in the opposite extreme when the polymer is wrung so
hard that it buckles and forms plectonemic
structures\cite{kamien}, which are stabilised by the finite thickness of
the polymer.
The transitional regime where the polymer
is neither straight\cite{sinha,nelson} nor plectonemic\cite{kamien} is
the subject of this paper.
Perturbation theory about the
straight line is not applicable and the simplifications that arise from
the energy dominated plectonemic regime do not obtain. We address this
regime by finding the minima of the energy functional in
the partition function.
These minima come in two families- the straight line family and the
``writhing" family, which deviates considerably from the straight line.
We study these configurations which
dominate the partition function in the transitional regime. We show that
these curves admit a local writhe formulation. This leads to a 
considerable simplification of the theory.

The paper is organized as follows.
Sec $II$ presents a recapitulation
of the geometrical notion of writhe of a simple space curve and how
it can be applied to understand DNA elasticity.
Sec $III$ deals with the mechanics of semiflexible polymers using
the classical mechanics of a top\cite{kirchoff}. We prove our main
result here:
the polymer configurations which dominate the partition function
admit a local writhe formulation. Brief remarks regarding
perturbations about the saddle point and stability are made in Sec
$IV$.
We end with some concluding remarks in Sec. $V$.

\section{Geometry}
\label{sec:1}
In this section we deal with the
essential geometric notions that determine the twist elastic properties
of linear molecules like DNA. This section is a recapitulation of some
material that has been presented earlier \cite{samsupabhi} and is being
summarised here for the reader's convenience. The experiments of Strick
{\it et al} \cite{strick}
are done on an open segment of DNA which is attached to a glass slide at 
one end and a
magnetic bead
on the other. The bead is pulled to stretch the molecule and turned to
``wind it up'' and induce supercoiling. For mathematical convenience, we
close the open segment of DNA by a reference ribbon (a ribbon is a
framed space curve) in a manner described
in \cite{samsupabhi} (see Figs. 4 and 5 of Ref\cite{fuller}
and references therein). The fixed
reference
loop (called the passive part)
is held fixed and our discussion concerns only the active part of the
curve. (It is also possible to choose the reference ribbon as
coming in from infinity along the ${\hat z}$ direction and going off
to infinity along the ${\hat z}$ direction. The precise choice of
reference does not matter.)

The basis of the analysis of DNA supercoiling is the celebrated
relation \cite{calugeranu,white,fuller1,fuller}:
\begin{equation}
Lk=Tw+Wr
\label{LktwWr}
\end{equation}
which relates the applied link (the number of times the bead is turned
in the experiment) to the twist (the twisting of the polymer about its
tangent vector) and the writhe (the bending of the tangent vector). The
problem neatly splits \cite{fuller,samsupabhi} into two parts: a
relatively
straightforward (local) description of
the twist elasticity, and the somewhat harder (nonlocal) treatment of
the writhe.

The writhe is a non-local quantity defined on closed simple
curves: Let the arc length parameter $s$ range over the entire length 
$L_0$ of the closed ribbon
(real ribbon $+$ reference ribbon) and
let us consider the curve ${\vec x}(s)$ to be a periodic function
of $s$ with period $L_0$.
Let ${\vec
R}(s,\sigma)={\vec x}(s+\sigma)-{\vec x}(s)$.
We write ${\hat t}(s)=\frac{d {\vec x}}{ds}$ for the unit tangent
vector to the curve.
Since the curve ${\vec x}(s)$ is simple,
${\vec R}(s,\sigma)$
is non-vanishing for $\sigma\neq 0,L_0$ and the
unit vector ${\hat R}(s,\sigma)$ is well-defined. It is easily checked
that as $\sigma \rightarrow \{0,L_0\},$
${\hat R} \rightarrow\{{\hat t},{-\hat
t}\}$
respectively. The C\u{a}lug\u{a}reanu-White  
writhe is given by\cite{dennis,calugeranu,white,fuller1,fuller}  
\begin{eqnarray}
{\cal W}_{CW}= \frac{1} {4\pi}\oint_{0}^{L_{0}}ds\int_{0+}^{L_{0}-}d\sigma
[\frac{d{\hat{R} (s,\sigma)}}{ds}
\times \frac{d{\hat{R}(s,\sigma)}}{d\sigma}]\cdot{\hat{R}.}
 \label{writhe}
 \end{eqnarray}
The non-locality of ${\cal W}_{CW}$ makes it difficult to handle
analytically.
However, the key point to note is that {\it variations} in ${\cal W}_{CW}$
are {\it local} \cite{fuller,fuller1}.
Let ${ {\cal C}}(\alpha)$ be a family of polymer
configurations parametrised by $\alpha$
of simple
closed curves with
writhe ${\cal W}_{CW}(\alpha)$.
Taking the ${\alpha}$ derivative of Eq. (\ref{writhe}) (which now
depends parametrically on $\alpha$) we find that the
resulting terms can be rearranged to give
\begin{eqnarray}
\frac{d{\cal W}_{CW}}{d\alpha} = \frac{1} {2\pi}\oint_{0}^{L_{0}}
ds [\frac{d{\hat{t} }}{ds}
\times \frac{ d{\hat{t}}}{d{\alpha}}]\cdot{\hat{t},}
\label{dwrithe}
\end{eqnarray}
 which clearly has the interpretation of the rate at which
${\hat t}(s, \alpha)$ sweeps out a solid angle in the
space of directions. Note that Eq.(\ref{dwrithe}) which
gives the
{\it change} in {\it writhe}
is a {\it single} integral \cite{bouchiat,fain} and therefore
a {\it local, additive}  quantity. This can be
expressed as a sum of the changes in the active
parts of the curve and therefore,
since the passive part is unchanged in the variation, the change
in writhe is entirely due to the active part $0\le s\le L$, where
$L$ is the contour length of the active part of the polymer.
Eq. (\ref{dwrithe}) can be rewritten as:
\begin{eqnarray}
\frac{d{\cal W}_{CW}}{d\alpha} = \frac{1} {2\pi}
\frac{d{\Omega}({\alpha})}{d\alpha}=\frac{1} {2\pi}\int_{0}^{L}
ds [{\hat{t}}
\times \frac{d{\hat{t} }}{ds}]\cdot\frac{ d{\hat{t}}}{d{\alpha}}
\label{dtwrithe}
\end{eqnarray}
where $\Omega$ is the solid angle
enclosed by the oriented curve \{${\hat t}(s)|0\leq s \leq L$\}
on the unit sphere of tangent directions
as $s$ goes from
$0$ to $L$ \cite{kamien}.
Note that $\Omega$ is
only defined modulo $4\pi$:
for a solid angle $\Omega$ to the left of the oriented curve ${\hat t}(s)$
is equivalent to a solid angle $(4\pi-\Omega)$ to the right.
$d\Omega/d\alpha$ is, however, well defined and local.
Integrating Eq. (\ref{dtwrithe}) we arrive at\cite{fuller,fuller1}:
\begin{eqnarray}
{\cal W}_{CW}({\alpha}) = \frac{1} {2\pi}
{\Omega} -1 +2n 
\label{twrithe}
\end{eqnarray}
where $n$ is an arbitrary integer. The constant of integration is fixed
by noting that the writhe of a simple planar curve vanishes  
(Eq.\ref{writhe}).

There {\it is} a quantity one can
construct from the writhe which is well defined on {\it all} curves
(not just simple ones)
$$w({ {\cal C}})=exp[i\pi{\cal W}_{CW}({ {\cal C}})]
= -exp[{i\Omega}/2] $$
is a complex number of modulus
unity which we call the
{\it wreathe}.
When a curve is passed through itself, ${\cal W}_{CW}$ jumps by $2$, but
the wreathe is unchanged. We can therefore smoothly extend
$w$ to all closed curves {\it including non-simple curves}.
Unlike the writhe, the wreathe
$w$ is a local quantity.
The wreathe can be exploited to define a new quantity,
the Fuller ``writhe'' 
\cite{fuller,fuller1} for a class of curves.
(Readers familiar with the geometric phase in quantum mechanics
\cite{Berry:1984jv,shapere,js,samsup} will recognise the relation
between the
wreathe, the
Fuller writhe, the Berry phase and the Pancharatnam connection.)
Let us fix a direction
(say $-{\hat z}$, which we call ``south'') and define ``south avoiding''
curves to be those for which the tangent vector never points south
\footnote{The sphere of tangent directions has perfect rotational
symmetry. The choice of the south direction $-\hat{z}$ is motivated
by the fact that in a real experiment one applies force and torque in
a specific direction which we choose to be along $+\hat{z}$.}.   
The Fuller ``writhe'' is defined for south avoiding curves.
Let us define the ``wreathe angular velocity''
${\cal A}_{\alpha} =
-iw^{-1}{\frac{dw}{d\alpha}}=1/2d\Omega/d\alpha$
on the family of curves ${\cal C}_{\alpha}$ parametrised by $\alpha$.   
Let us choose a fiducial curve ${{\cal C}_*}$, of length $L$, whose
tangent vector is identically north pointing. We take this as a standard
for the ``active part'' of the curve, the ``passive part'', of course
being held fixed in the discussion.
Observe that all south avoiding curves are deformable to the fiducial
curve
${{\cal C}_*}$. One simply deforms the tangent vector ${\hat t}(s)$
along the unique shorter geodesic connecting ${\hat t}(s)$
to the north pole. We now define the ``Fuller writhe'' as
$${\cal W}_F=1/\pi \int_{{\cal C}_*}^{ {\cal C}} d\alpha {\cal
A}_\alpha-1.$$
Writing the unit tangent vector as ${\hat t}=(\sin{\theta}\cos{\phi},   
\sin{\theta}\sin{\phi},\cos{\theta})$,
$\int_0^{1}{d{\alpha} \frac{d\Omega}{d\alpha}}$ can be written as
$\int{ds\frac{d{\phi}}{ds}(1-\cos{\theta})}$ for all curves for which the
tangent vector never points towards
the south pole of the sphere of tangent directions. We
can therefore write
\cite{bouchiat,fain}
a local writhe formula on such curves which we call ``south avoiding
curves'':
\begin{eqnarray}
{\cal W}_{F} = {\frac{1}{2\pi}}\int{ ds {(1- \cos{\theta})
{\frac{d{\phi}}{ds}}}} -1.
\label{sangle}
\end{eqnarray}
While (Eq. \ref{sangle}) is expressed in local co-ordinates on the   
sphere,
it has a clear geometric meaning:
$2\pi(1+{\cal W}_F)$ is equal to the solid angle swept out by the unique 
shorter
geodesic connecting the tangent
vector ${\hat t}$ to the north pole.
This definition is explicitly {\it not} rotationally
invariant, since it uses a fixed fiducial
curve ${{\cal C}_*}$ and singles out a preferred direction.

To summarise, ${\cal W}_{CW}$ is defined on all self-avoiding curves,
${\cal W}_F$ on
all south avoiding curves.
When a curve passes through itself,
${\cal W}_{CW}$ jumps by two and when the tangent vector to a curve swings
through the south pole, ${\cal W}_F$ jumps by two units. The writhe is a 
real
number which has {\it both}
geometric and topological information. The topological part is the
integer part of $\frac{{\cal W}_{CW}}{2}$ and the geometric part is the
fractional part of $\frac{{\cal W}_{CW}}{2}$. The geometric part is
completely
captured by wreathe but the topology is lost since wreathe
is insensitive to changes in writhe by $2$ units.
From the definitions it is clear that
on a family of curves ${\cal C}_{\alpha}$ parametrised by $\alpha$,
\begin{equation}
-i{\frac{{w}^{-1}}{\pi}}{\frac{dw}{d\alpha}}=
\frac{d{\cal W}_{CW}}{d\alpha}=
\frac
{d{\cal W}_{F}}{d\alpha}
\label{change}
\end{equation}
So, continuous changes in writhe are the same whether measured by ${\cal
W}_{CW},{\cal W}_F$
or $w$. We have adjusted the reference ribbon and the
definition of Fuller writhe so that these two notions of writhe
${\cal W}_{CW}$ and ${\cal W}_F$ agree on
the reference curve. Note that $w$ is well-defined on {\it all} curves.
For a closed circuit of curves $\{{\cal
C}_{\alpha},(0\le\alpha\le1),{\cal C}_0={\cal C}_1\}$, we have the
``closed
circuit theorem'' \cite{samsupabhi}: In a closed circuit of curves,
the total number of south crossings and self crossings
(both counted with sign) are the same. This follows easily by
integrating (Eq. \ref{change}) along a closed circuit.

This definition of ``Fuller writhe'' is motivated by a Theorem of
Fuller\cite{fuller,fuller1}, which uses a reference curve and states   
precise
conditions (which are frequently misunderstood\cite{neukirch}) under
which the writhe difference between the two curves
can be written as a local integral.
Under deformations of the reference curve which are {\it both south
avoiding
and self avoiding} (we follow \cite{neukirch} in calling these ``good
deformations''), Eq. (\ref{change}) implies that the equality ${\cal
W}_{CW}={\cal W}_F$ is maintained.
These deformations are
the ones which satisfy the conditions of Fuller's theorem
[Eq. 6.4 in Ref. \cite{fuller}]. We refer to curves which can be
reached from the reference curve
by ``good deformations'' as ``good curves''.

 To avoid misunderstanding, we remark that the ``Fuller writhe'' is a
distinct quantity from ${\cal W}_{CW}$ and
is {\it not} equal to the C\u{a}lug\u{a}reanu-White writhe.
This is obvious since the first is local, while the second is not.
Strictly speaking Fuller only defines a local writhe on ``good
curves'', where it is equal to the CW writhe. We find it 
convenient\cite{samsupabhi} to extend Fuller's defintion to
all south avoiding curves and not just ``good curves''.
On ``good curves'', the two notions of writhe agree and we have ${\cal  
W}_F$ {\it equal to} ${\cal W}_{CW}$,
but the two quantities are different in general.
The main claim of \cite{samsupabhi} is that ``good curves'' dominate the
partition function in a region of the parameter space.
We will substantiate this claim in the next section
by identifying the dominant configurations that contribute to the
partition function in a saddle point approximation and showing that
they are ``good curves''.

In the high force region\cite{sinha,nelson} this point is obvious since
the polymer fluctuates about the straight line configuration which is a 
good
curve. The content of the following section is that even at
lower forces (or at large torques, close to buckling) when the polymer 
is far from straight, the minima 
of the energy are ``good curves''.  The analysis involves a combination
of mechanics, geometry and topology.

\section{Mechanics}
\label{sec:2}
Semiflexible biopolymers like DNA are subject to thermal fluctuations
in a
cellular environment. Therefore the natural theoretical framework for   
studying the elastic properties of such systems is statistical
mechanics,
which involves a competition between energy and entropy.
In the experiment, one controls the force $F$ on the bead and the link
$Lk$, the number of times the bead is turned.
This brings up the theoretical problem of computing the partition
function $Z(F, Lk)$, the number of configurations (counted with Boltzmann
weight) of the ribbon which have a given link (the link distribution).
We know that the link distribution is a
convolution of the writhe distribution
and the twist distribution\cite{bouchiat,sinha}.
The latter is a simple Gaussian integral, which is easily computed.
The real problem is to compute the writhe distribution of an
inextensible curve.
By simple
transformations \cite{samsupabhi} the problem of determining the 
link distribution of an inextensible ribbon can be reduced to computing  
the
writhe distribution of a space curve.  
The partition function to be computed is
\begin{equation}
Z(F,W) = \sum_{\it {\cal C}}{exp[-{\cal E}({\cal
C})/k_BT]\delta({\cal W}_{CW}-W)}
\label{partfnwlrc} 
\end{equation}
where $F$ is the force, ${\cal E}({\cal C})$ the energy
(defined below in eq.(\ref{energy}))
and ${\cal W}_{CW}$ is
the writhe. The sum is over all allowed configurations of the
polymer\cite{samsupabhi}. We expect the sum to
be dominated by configurations which minimise the energy
functional.
Configurations near the
minimum will also contribute due to
thermal fluctuations.
We first do a purely classical elastic analysis taking only the
energy into account, while
ignoring the entropy\cite{fuller1,love,maddockspnas,fain}. This
allows us to draw on physical intuition derived from the elasticity of 
beams,
cables, telephone cords and ribbons and paves the way for a fuller
treatment which incorporates thermal fluctuations around the classical
solutions. 
We analyze the classical elasticity of a torsionally
constrained stretched semiflexible polymer.

The problem can be reduced to minimizing the energy (we use a dot for
the $s$ derivative, it reinforces the analogy \cite{kirchoff} with the
dynamics of tops)
\begin{equation}
 {\cal E}[\hat{t}(s)] = \frac{1}{2} \int_0^L \dot{\hat t}.\dot{\hat t} ds 
- \int_{0}^{L} \vec{F}.\hat{t} ds
\label{energy}
\end{equation}
of a space curve $\vec{x}(s)$ whose tangent vector is $\hat{t}(s) =
\frac{d\vec{x}}{ds}$,
subject to a writhe constraint
\begin{equation}
{\cal W}_{CW}=W
\label{writheconstraint}
\end{equation}
where, ${\cal W}_{CW}$, the C\u{a}lug\u{a}reanu-White
writhe has been defined earlier. The
tangent vector $\hat{t}(s)$ is varied subject to the boundary conditions
$\hat{t}(0) = \hat{t}(L) = \hat{z}$
fixing the tangent vector to the curve at both ends.
As mentioned clearly in \cite{nelson,sinha,samsupabhi}, the full
problem involving the link distribution (with finite twist elasticity) can
be reduced to computing the writhe distribution,
which is essentially the limit of infinite twist rigidity
($C\rightarrow\infty$).

Using the method of Lagrange multipliers we arrive at
\begin{eqnarray}
\delta {\cal E}({\cal C}) = 2 \pi \tau\delta {\cal W}_{CW}
\label{vary}
\end{eqnarray}
where $2\pi\tau$ is a Lagrange multiplier. $\tau$ can be physically
interpreted as a torque.
The variation in the energy $\delta{\cal E}$ is given by 
\begin{eqnarray}
\delta {\cal E} = - \int_{0}^{L} (\ddot{\hat t} + \vec{F}) \cdot 
\delta{\hat t}
\label{varya}
\end{eqnarray}  
The variation of ${\cal W}_{CW}$ (as explained earlier
(\ref{dtwrithe}))
is a local quantity
\begin{eqnarray}
\delta {\cal W}_{CW} = \frac{1}{2\pi}\int_{0}^{L} [{\hat t} \times 
\dot{\hat t}]\cdot \delta{\hat t} ds
\label{d2writhe}
\end{eqnarray}  
and the Euler-Lagrange (E-L) equations are
\begin{eqnarray}
-\ddot{\hat t} -
{\vec F}=\tau({\hat t}\times{\dot{\hat t}})- \gamma{\hat t},
\label{eqnofmot} 
\end{eqnarray}
where the term $\gamma{\hat t}$ arises since $\delta{\hat t}\cdot{\hat
t} = 0$. Note that the E-L equations are local differential equations
rather than the nonlocal integral equations that one may have
expected from the variation of a non local quantity ${\cal W}_{CW}$.
Thus, the {\it variations} of writhe ${\cal W}$ are {\it local}. As we
will
see, this is the {\it key idea} that allows us to prove that ``good
curves''
dominate the partition function.

The {\it statics} of a twisted beam (or cable or DNA)
is formally similar to the
{\it dynamics}
of a heavy symmetrical top, a fact that has been
well known since Kirchoff\cite{kirchoff}. The analogy is
useful for integrating the Euler-Lagrange
equations. We use quotes for the analogous top quantities. The ``kinetic
energy'' is given by $T=\frac{1}{2}{\dot{\hat t}} \cdot \dot{\hat t}$
and the ``potential energy'' is 
$V={\vec F}\cdot {\hat t}$.
The total ``energy''
\begin{eqnarray}
{\cal H} = T + V =\frac{1}{2} 
(\dot{\theta}^2+\sin^{2}\theta\dot{\phi}^2)+F\cos\theta
\label{hamex}
\end{eqnarray}
is a ``constant of the motion'' as is the $z$ component of the ``angular
momentum''
\begin{eqnarray}
J_{z}=({\hat t} \times\dot{\hat t})_z-\tau{\hat t}_{z} =
\sin^{2}\theta\dot{\phi}   
- \tau\cos\theta,  
\label{zang}
\end{eqnarray}
where we have introduced the usual polar coordinates on the space of
tangent vectors ${\hat
t}=(\sin{\theta}\cos{\phi},\sin{\theta}\sin{\phi},\cos{\theta})$.
Using these
``constants of the motion'' we
reduce
the
problem to quadratures as described in \cite{goldstein}. The basic
equations are
\begin{eqnarray}
\frac{\dot\theta^{2}}{2} = {\cal H} - F \cos\theta -
\frac{(J_{z} + \tau\cos\theta)^{2}}{2\sin^{2}\theta}
\label{hamex1}
\end{eqnarray}
\begin{eqnarray}
\dot{\phi} = \frac{J_{z} + \tau\cos\theta}{\sin^{2}\theta}.
\label{phidot}
\end{eqnarray}
Setting $u = \cos\theta$, we find that
\begin{eqnarray}
{\dot u}^2= f(u)
\label{solu}
\end{eqnarray}
where $f(u)$ is a cubic polynomial in $u$.
Our boundary conditions imply that $\theta = 0$ is a point on the
solution. Since ${\dot \theta}^{2}$
has  to be finite at $\theta = 0$, we have $J_{z} = -\tau$. The form of
$f(u)$ simplifies to
\begin{equation}
f(u)=(1-u)[2({\cal H}-Fu)(1 + u)  - \tau^{2} (1-u)].
\label{fofu}
\end{equation}
(\ref{phidot}) can be rewritten as
\begin{eqnarray}
\dot{\phi} = \frac{-\tau}{1+u}.
\label{phidotnew}
\end{eqnarray}
From (\ref{hamex1}) we also find that ${\cal H}\ge F$, else
${\dot \theta}^{2}$ would have to be negative at $\theta=0$.  We see
that $f'(u)|_{u=1}=-4({\cal H}-F)\le 0 $ at $u=1$. At $u=-1$,
$f(-1)=-4\tau^2\le0$ and so $f(u)$ has one physical root at $u=1$ and
the other at $u_0$ between $-1$ and $1$: $-1\le u_0\le1$. The third root
$u_1$ is generally (except for $u_1=\pm1$) outside
the physical range and $u_1>1$ for positive force and $u_1<-1$ for
negative force. Since $f(u_0)=0$, we find from (\ref{fofu}) the useful
relation
\begin{equation}
\tau^2=\frac{2({\cal H}-Fu_0)(1+u_0)}{1-u_0}
\label{useful}
\end{equation}
The ``motion'' is confined between
the 
turning points  $1$ and $u_0$: $u_0\le u\le1$. We write
$u_0=\cos{\theta_0}$, where $\theta_0$ is the turning point for   
$\theta$.
Just as in the top problem, we find that the solution $u(s)$ is
periodic with a period given by
\begin{equation}
P = 2\int^{1}_{u_{0}}
\frac{du}{\sqrt{f(u)}}.
\label{period}
\end{equation}
Because of the boundary conditions, we have $L=nP$, where $L$ is the
length of the polymer and $n$ is an integer. $L$ and $n$ fix the  
integration constant ${\cal H}$. We will see in the next section
that solutions with $n>1$ are not minima of the energy.
We therefore restrict ourselves to $n=1$. Our interest is solely
in stable solutions, that is local minima of the energy.

General solutions to these equations can be found in terms of elliptic
functions but a pedestrian approach gives more immediate insight.
The simplest solution is the straight line $u(s)=1$ for all $s$, which
solves the E-L equations (\ref{eqnofmot}) for all $F,\tau$ with $F={\cal
H}$ for any $L$. However, the straight line cannot accommodate writhe
and we have to look at the ``writhing family'' of solutions parametrised 
by
$F,\tau$. These start from $u(0)=1$ at $s=0$, and have a turning point
at $s=s_0$ ($u(s_0)=u_0$) and return to $u=1$ at $P=2s_0=L$. 
These solutions can be explicitly written down in terms of elliptic 
functions and vary continuously with $\tau$ in the allowed range 
$|\tau|<\sqrt{4F}$.
Integrating
(Eqs. \ref{hamex1},\ref{phidotnew}) gives $(u(s),\phi(s))$ which gives
us
the solution to the E-L equations ${\hat t}(s)$.

From the ``top'' point of view, ${\hat t}(s)$ {\it is} the solution. 
However, in polymer physics, ${\hat t}(s)$ is only a convenient way
of describing the polymer configuration ${\vec x}(s)$, which is given
by
\begin{equation}
{\vec x}(s)=\int_0^{s} ds' {\hat t}(s')
\label{xofs}  
\end{equation}
The use of ${\hat t}(s)$ rather than ${\vec x}(s)$,   
in the variational problem (Eq. \ref{energy}) leads to a considerable
simplification,
since the equations of motion are second order rather than fourth order.
However, there is a price to be paid. After solving for the ``tantrix''
${\hat t}(s)$ (plotted in a special case in Fig. 1), we still have to
integrate (Eq. \ref{xofs}) to produce
the actual polymer configuration ${\vec x}(s)$, (Fig. 2)
in real space.
\begin{figure}
\includegraphics[height=8.7cm,width=8.7cm]{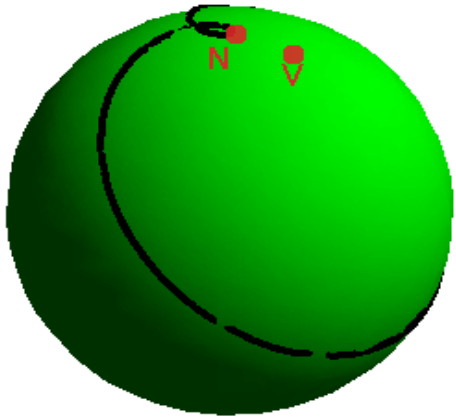}
\caption{The black curve (tangent indicatrix)
traced on the unit sphere (green
online) by the unit tangent vector to a
long polymer for a force $F=.11$  and torque $\tau=.4$.
The dots (in red online) indicate the tips of the unit vectors ${\vec
V}$ (marked V) and  ${\vec z}$ (marked N for North).}
\end{figure}
\begin{figure}
\includegraphics[height=7.0cm,width=3cm]{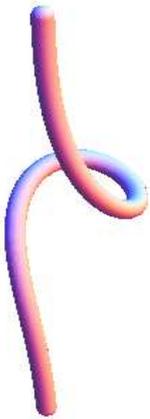}
\caption{Figure shows the polymer configuration in real space for the
same parameters as Fig. 1.
The curve has been thickened, shaded and truncated for easy three
dimensional visualisation.}
\end{figure}
In the polymer context it is meaningful to ask:
does this configuration intersect itself? Such a question is
not significant for tops. Nevertheless, as we will
see, the motion of a heavy symmetrical top helps us answer this
question.

We prove the main result of this paper: all stable solutions
of the E-L equations are ``good curves''in the sense of the last section.
For the straight
line family, the result is
obviously true since the straight line {\it is} the reference curve.
The effort here is showing that all stable members of the writhing
family
$\hat{t}(s)$
parametrised by $(F, \tau)$ are ``good curves''.
To do this we prove that the writhing family is nowhere self
intersecting or south pointing for $-1< u_0\leq 1$.
Since $u_0 = 1 $ is the straight line ${\hat t}(s)={\hat z}$,
which we take as the reference curve,
this proves that the writhing family can be
deformed (via a deformation of the parameter $u_0$) to the reference
curve by ``good''
deformations
and the result follows. Below, we explicitly exclude the case $u_0=-1$,
which will be treated separately in the next section.

It is evident that the writhing family is nowhere south pointing for
$-1< u_0\leq 1$ since the turning point $\theta_{0} < \pi$ and the
tangent vector ${\hat t}(s)$ never points to the south pole. To prove
that the writhing family is
nowhere self intersecting requires considerably more work. The proof
consists of finding a direction ${\vec V}$ along which the tangent
vector ${\hat t}$ to the polymer has positive component. This
implies that the polymer never bends back to intersect itself.
The proof uses the E-L equations in an essential way but does not
rely on the explicit form of the solution.

We first orient our
axes by rotating about the  $z$
axis so that $\phi(s_0)=0$ and the tangent vector
${\hat t}(s)=(\sin{\theta(s)}\cos{\phi(s)},
\sin{\theta(s)}\sin{\phi(s)},\cos{\theta(s)})$
lies in the $x-z$ plane at the turning point.
Consider for each $u_0=\cos{\theta_0}$ the constant ($s$
independent) unit vector
$\vec{V} = (-\cos\theta_{0}, 0,\sin\theta_{0})$.
and the function
 \[
g(s) = {\hat t}(s)\cdot \vec{V}.
\]
Note that the function $g(s)$  is symmetric\footnote{
This follows since $\theta(s)$ is an even function of $s-s_0$ and 
$\phi(s)$ 
is an odd function of $s-s_0$. (see eq.(\ref{phidotnew})). Briefly, the 
orbit is symmetric about its turning point $s_0$.} about $s=s_{0}$
within the period $0\le s\le 2s_0=P=L$.
We show that a) either $g(s)$ vanishes identically
or b) $g(s)>0$ for all $s\ne s_0$. We need only consider
the range $ s_{0}\le s < 2s_{0}$, so that $\theta(s)>0,u(s)<1$.
$g(2s_0)$ is fixed by boundary conditions to be $\sin{\theta_0}\ge0$.
The positivity of $g(s)$ immediately implies our central result (see
below):
The polymer configurations under stretch and torque are such that
the tangent vectors to the polymer all lie in one hemisphere
of the sphere of tangent directions.

The proof proceeds in several steps.\\
{\it Step 1:} $g(s)$ vanishes at the turning point $s_0$ and has a local
minimum
at $s_0$ if it does not vanish identically.\\
Proof: 
Remembering that ${\dot \theta}=0$ at the turning point, we find 
\begin{equation}
{\hat t}(s_0)=(\sin{\theta_0},0,\cos{\theta_0})
\label{tform1}
\end{equation}
and
\begin{equation}
{\dot{\hat t}}(s_0)=(0,\sin{\theta_0}{\dot \phi},0)
\label{tform2}
\end{equation}
from which follows
\begin{eqnarray}
g(s_0)={\hat t}(s_0).{\vec V}=0\\
{\dot g}(s_0)={\dot {\hat t}}(s_0).{\vec V}=0
\label{gform} 
\end{eqnarray}
So $s_0$ is a stationary point of $g(s)$. To show that $g(s)$ has a
local
minimum, we compute ${\ddot g(s)}$ using the E-L
equations (\ref{eqnofmot}) and
(\ref{phidotnew},\ref{tform1},\ref{tform2}) to
find
\begin{equation}
{\ddot g(s_0)}=\sin{\theta_0}(\frac{\tau^2}{1+u_0}-F)
\label{gddot} 
\end{equation}
Substituting for $\tau^2$ (\ref{useful}) and simplifying
gives
\begin{equation}
{\ddot g(s_0)}=\frac{\sin{\theta_0}}{1-u_0}(2{\cal H}-F(1+u_0))\ge0
\label{gddot2}
\end{equation}
since ${\cal H}\ge F$.
The equality occurs only if $\sin{\theta_0}$ vanishes ($u_0=\pm1$)
in which case $g(s)$ vanishes identically. $u_0=1$ describes a straight
line segment, which is the reference curve and therefore a ``good
curve''. $u_0=-1$ has been excluded from consideration. 
Below we will exclude both cases ($u_0=\pm1$) in
which $g(s)$ vanishes identically.
This proves that $s_0$ is a local {\it minimum} of $g(s)$ for
($u_0\ne\pm1$).

{\it Step2:} At stationary points of $g(s)$,
\begin{equation}
g(s) u(s) >0
\label{sign}
\end{equation}
for $s\ne s_0$.

{\it Proof of Step 2:}
Since $\dot{g}(s)=\dot{\hat{t}}.{\vec V}=0$ at a stationary
point, $\dot{\hat{t}}$ is orthogonal to both ${\vec V}$ and $\hat t$ and
can be written. $\dot{\hat{t}}=A {\vec V}\times {\hat t}$ for some
nonzero $A$.
 ``Angular momentum'' and ``energy'' conservation give us
\begin{eqnarray}
\dot{\hat t}. \dot{\hat t} = 2({\cal H}-Fu)
\label{encons}
\end{eqnarray}
\begin{eqnarray}
({\hat t} \times \dot{\hat t})_{z} = -\tau (1-u),
\label{Jcons} 
\end{eqnarray}
which can be rewritten as
\begin{eqnarray}
|A|^{2} (1 -g^2) =2({\cal H}-Fu)
\label{encons2}
\end{eqnarray}
and
\begin{eqnarray}
A(\sin\theta_{0} - gu ) = -\tau (1-u)
\label{Jcons2}
\end{eqnarray}
In Eq.(\ref{encons2}) both sides are positive since $u<1$.
Squaring (\ref{Jcons2}) and
eliminating $|A|^{2}$ and $\tau^2$ between
(\ref{useful},\ref{encons2},\ref{Jcons2}), we find that at
a stationary point of $g(s)$, $g(s)$ and $u(s)$ satisfy
\begin{equation}
(\sin\theta_{0} -gu)^{2} < \frac{({\cal H}-Fu_0)(1+u_0)}{(1-u_0)} X(u)
\label{iforget}
\end{equation}
where $X(u)=(1-u)^2/({\cal H}-Fu)$ is a strictly decreasing function of
$u$ in the range of interest. From $X(u)<X(u_0)$ for $u>u_0$
it follows that
\begin{equation}
(\sin\theta_{0} -gu)^{2} <(\sin\theta_{0})^{2}
\label{uforget}
\end{equation}  
which implies (\ref{sign}).

{\it Step3:} $s_0$ is a global minimum of $g(s)$: $g(s)>0$ for
$s\ne s_0$.

Proof: note that since $s_0$ is a local minimum of $g(s)$ (from
Step 1), as $s$ increases from $s_{0}$, $g(s)$ {\it increases} from 0
(since $s=s_{0}$ is a {\it minimum}). If there are no more stationary
points $g(s)$ remains positive and we are through. If the
next stationary point occurs at $s_1>s_0$, $\dot{g}(s)|_{s=s_{1}}=0$
we have $g(s_{1}) > 0$. It follows from Step 2 that
 $u(s_{1}) > 0$.
Since $u(s)$ is a  monotonic function of $s$ in the range $(s_0\le s<
2s_0)$, we have
$u(s) > 0$
for $s > s_{1}$ in this range. All minima of $g(s)$ for $s > s_1$,
therefore have
$g(s) > 0$ (again using Step 2). It
follows that $g(s) > 0$ for all $s$ (except the one point $s_{0}$,
where $g(s_{0})=0)$. From the symmetry of $g(s)$ it
follows that $g(s)$ is positive everywhere except at the turning point
($s_0$,  where it vanishes).

This immediately implies our main result since if
any of the minimum energy configurations intersect themselves,
$\vec{x}(s_{1}) = \vec{x}(s_{2})$ for $s_{2} > s_{1}$, we have
\[
\int^{s_{2}}_{s_{1}} {\hat t}(s)\cdot ds = 0
\]
contracting this equation with $\vec{V}$ gives us
\[
\int^{s_{2}}_{s_{1}} g(s) ds= 0
\]
which is a contradiction since $g(s) > 0$ for all $s\neq s_{0}$.
We have thus proved that the writhing family does not intersect itself
for $-1< u_0\leq 1$. Since $u_0$ can be continuously deformed to
$1$ along the writhing family, we conclude that the family consists
of ``good curves'' which admit a local writhe formulation. We now
consider variations around the solutions of the E-L equations
to test for stability.

\section{Fluctuations around the saddle point}
In a saddle point approximation to the partition function the dominant
thermodynamic contributions are expected to arise from the global
minimum of the energy. We can also
expect contributions from local minima due to ``metastable'' states,
when the time scale of the experiment is not large enough to find the
true minimum. All minima, local and global will satisfy the E-L equations
and so will be among the solutions described in the
last section. However, not all solutions to the E-L equations are minima
of the energy. There are some configurations which, if perturbed, will
go to nearby states of lower energy and decay.
These configurations do not contribute significantly to the partition 
function.
To weed these out we need to consider the stability of solutions by
perturbing the classical solutions to see if there are neighboring
lower energy configurations. (We do not concern ourselves
here with zero modes arising from symmetry breaking. They contribute
an innocuous multiplicative constant to the partition function.)

For example, we note that solutions with $L=nP$ and $n>1$ are unstable.
This is seen by explicitly constructing neighboring configurations
of lower energy.
Suppose $n > 1$ and consider the configuration
\{\(u_{\rm old}(s),\varphi_{\rm old}(s)\)\} which solves
the Euler Lagrange equations.
Construct a new configuration
\begin{equation}
\left.\begin{array}{cc} u_{\rm new}(s) &= u_{\rm old}(s) \\  \varphi_{\rm 
new}
(s) &= \varphi_{\rm old}(s)
\end{array}\right\}{\rm for}\; s \leq P
\label{newconfig}
\end{equation}
\begin{equation}
\left.\begin{array}{cc} u_{\rm new}(s) &= u_{\rm old}(s) \\
\varphi_{\rm new}(s) &= \varphi_{\rm old}(s) + \alpha   
\end{array}\right.\;\;{\rm for}\; s \geq P
\end{equation}
where $\alpha$ may be an arbitrarily small real number. The new
configuration
\begin{enumerate}
\item satisfies the boundary conditions,
\item has the same energy as the old one (since the energy is a local
integral),
\item has the same writhe as the old one (since changes of 
writhe are local integrals),
\item is {\it not} a solution to the
${\cal E}L$ equations at $s=P$ (because of the break in the curve
at this point).
\end{enumerate}
It follows that there are neighboring configurations
with lower energy than the old configuration and the same writhe.
We conclude that the solutions with $n>1$ are not local minima of the
energy.   
A similar remark applies to the straight line family if the torque
exceeds a critical torque $\tau_c$ given by
$(\tau_c/2)^2=F+(\frac{\pi}{L})^2$, 
the classical condition for torsional buckling of a rod.
This can be directly seen by perturbing the straight line.
For $\tau>\tau_c$ there are {\it no} stable solutions of the Euler
Lagrange equations and the polymer must double back on itself
and enter the plectonemic regime.

We now turn to a discussion of the exceptional member of the writhing
family $(u_0=-1)$ characterised by $\tau=0$.
Note that as $\tau$ ranges from $-\tau_c$ to $\tau_c$, the writhing
family goes in a closed circuit from the straight line to the straight
line. The closed circuit has one south crossing (at $\tau=0$ or
equivalently $u_0=-1$).
From the ``closed circuit theorem'' \cite{samsupabhi}, it
follows that the circuit must also have one self crossing. Since none
of the other members of the writhing family have self intersections,
it follows that self crossing must occur at $\tau=0$. Thus the closed
circuit has one self crossing and one south crossing,
both occuring at the same value of the parameter $\tau=0$.
The writhing family curve $\tau=0$ does not belong to the configuration
space of either self avoiding or south avoiding models and both ${\cal
W}_{CW}$ and ${\cal W}_F$ are ill defined. One can expand the  
configuration space to allow all curves and use a formulation in which
energy is minimised for fixed {\it wreathe}.
It can then be shown by an argument
very similar to that given above (\ref{newconfig}) that the curve is
unstable. By rotating
the second half of the curve by a small angle $\alpha$ relative to the
first, keeping the energy and wreathe the same, we can produce
neighboring curves which are not solutions of the E-L equations and
can therefore be perturbed to lower their energy.

The writhing family curves in the regime $0<|\tau|<\tau_c$,
include stable
solutions, and we can take into account thermal fluctuations in a
Gaussian approximation about the local
minimum of energy. Since any stable curve of the
writhing family is at a finite distance and energy from south and
self intersection, we
can expect
small fluctuations around the original curve to be  `` good curves''.
Large fluctuations may
lead to ``bad curves'', but their contributions to the partition
function will be exponentially suppressed by
the Boltzmann weight,
as made clear by the example of \cite{comment}, and
not affect the {\it statistical} predictions of the model.
Thus the writhing family can be used as a base for a clean discussion of
thermal fluctuations without getting embroiled in the topological
subtleties of writhe.
Such a treatment goes far beyond perturbation
theory around a straight line since the writhing family is far from
straight.

To summarize, our analysis shows that there
is a range of force and torque relevant to single molecule experiments
for which the polymer configuration curves are ``good'' curves.

\section{Conclusion} 

In this paper we present a classical mechanical approach to
biopolymer elasticity based on
well known analogies between twisted polymers and tops. Our main result
is that the dominant configurations in the partition function  are
``good curves'' in a topological sense.
This result permits a local formulation of the writhe  and a   
non perturbative attack on the problem of twist elasticity by
incorporating thermal fluctuations around the writhing family.
Such an approach is expected to work well in the energy dominated
regime, when the polymer is twisted and stretched.
In this paper, we have restricted ourselves to  the
mathematical and conceptual issues regarding the use of writhe.
Our main result is useful in further developing the 
theory\cite{papb}.
Applying these ideas to the WLC leads to detailed predictions about the 
statistical mechanical
behaviour of twisted stretched polymers. These will be described  
elsewhere.

The experiments of Strick et al \cite{strick} have been theoretically
analysed by Nelson \cite{nelson} and Bouchiat and Mezard
\cite{bouchiat}. The analysis of Nelson is perturbative about the high
force limit (see also \cite{sinha}) and not controversial. However,
Bouchiat and Mezard \cite{bouchiat} ventured into the non-perturbative
regime of low forces. By using a local writhe formula they were able
to achieve agreement with experiments. However, this step is open to
criticism\cite{maggs}. In fact, the ``derivation'' \cite{bouchiat} of 
a ``local writhe formula'' is flawed since it is based on the wrong 
assumption that the Euler angles are continuous.
In \cite{samsupabhi}, it was suggested that in an
approximation, the success of Bouchiat and Mezard's model could
be due to the fact that the partition function was dominated by ``good
curves''. The present work establishes unambiguously that this is indeed
the case. In \cite{comment} we have argued that, in the high Link regime,
just as a self-avoiding polymer winds around itself forming plectonemes,
forming ``australonemes''. The energies of these two configurations are
similar, although the configurations are not. As a result we find that
\begin{equation}
Z_{F}({\vec r},Lk)\approx Z_{CW}({\vec r},Lk)
\label{approx}
\end{equation}
i.e. the partition functions based on local and non-local formulations  
of writhe are approximately equal to each other for a wider range of
parameters than one may have naively expected. The parameter space
consists of
$(F,\tau,L)$ and the energy dominated regime includes short
polymers ($L$ of the order of persistence length), stretched
polymers($F$ somewhat greater than $\tau^2/4$) and wrung
(or plectonemic) polymers ($\tau^2/4$ somewhat greater than $F$).

We expect this work to be of interest to researchers in this
interdisciplinary field of biopolymer elasticity. More work is
necessary  to investigate the transition from the
energy dominated regime
to the entropy dominated regime. Analytic methods,
numerical experiments and real experiments,
all have a role to play in this endeavour.
Our mathematical analysis of this problem
relies on insights gained by Fuller \cite{fuller,fuller1}
and uses them to understand the entropic elasticity
of twisted polymers.


%
%

\begin{acknowledgements}
It is a pleasure to thank Abhishek Dhar, A. Jayakumar,
Anupam Kundu  and Sanjib Sabhapandit for their comments on our
manuscript.
\end{acknowledgements}



%
%
\end{document}